\documentclass[10pt]{article}
\usepackage{geometry} 
\usepackage{amsmath}
\usepackage{amssymb}
\usepackage{amsthm}
\usepackage{graphicx}
\usepackage{chicago}
\usepackage{url}
\usepackage{multirow}

\newcommand{\Ebb}{\ensuremath{\mathbb{E}}}

\newcommand{\lgd}{\ensuremath{{L_{GD}}}}

\newcommand{\CVA}{\ensuremath{\text{CVA}}}
\newcommand{\FVA}{\ensuremath{\text{FVA}}}
\newcommand{\ColVA}{\ensuremath{\text{ColVA}}}
\newcommand{\MVA}{\ensuremath{\text{MVA}}}
\newcommand{\KVA}{\ensuremath{\text{KVA}}}

\newcommand{\Vhat}{\ensuremath{\widehat{V}}}
\newcommand{\VhatBoth}{\ensuremath{\Vhat_\text{C+H}}}
\newcommand{\VhatC}{\ensuremath{\Vhat_C}}
\newcommand{\VhatH}{\ensuremath{\Vhat_H}}
\newcommand{\Vhati}{\ensuremath{\Vhat_i}}

\newcommand{\dVHdt}{\frac{\partial \Vhat_H}{\partial t}}
\newcommand{\dVHdP}{\frac{\partial \Vhat_H}{\partial P_r}}
\newcommand{\dVVHdPP}{\frac{\partial^2 \Vhat_H}{\partial P_r^2}}

\newcommand{\dUHdt}{\frac{\partial U_H}{\partial t}}
\newcommand{\dUHdP}{\frac{\partial U_H}{\partial P_r}}
\newcommand{\dUUHdPP}{\frac{\partial^2 U_H}{\partial P_r^2}}

\newcommand{\half}{\frac{1}{2}}

\title{Behavioural effects on XVA}
\author{Chris Kenyon and Hayato Iida\footnote{Contacts: chris.kenyon@mufgsecurities.com,  hayato.iida@mufgsecurities.com.  This paper is a personal view and does not represent the views of MUFG Securities EMA plc (“MUSE”).  This paper is not advice.  Certain information contained in this presentation has been obtained or derived from third party sources and such information is believed to be correct and reliable but has not been independently verified.  Furthermore the information may not be current due to, among other things, changes in the financial markets or economic environment.  No obligation is accepted to update any such information contained in this presentation.  MUSE shall not be liable in any manner whatsoever for any consequences or loss (including but not limited to any direct, indirect or consequential loss, loss of profits and damages) arising from any reliance on or usage of this presentation and accepts no legal responsibility to any party who directly or indirectly receives this material.}}
\date{09 Mar 2018\\\vskip5mm Version 1.00} 

\begin{document}
\maketitle
\begin{abstract}
Bank behaviour is important for pricing XVA because it links different counterparties and thus breaks the usual XVA pricing assumption of counterparty independence.  Consider a typical case of a bank hedging a client trade via a CCP.  On client default the hedge (effects) will be removed (rebalanced).  On the other hand, if the hedge counterparty defaults the hedge will be replaced.  Thus if the hedge required initial margin then the default probability driving MVA is from the client not from the hedge counterparty.  This is the opposite of usual assumptions where counterparty XVAs are computed independent of each other.  Replacement of the hedge counterparty means multiple CVA costs on the hedge side need inclusion.   Since hedge trades are generally at riskless mid (or worse) these costs are paid on the client side, and must be calculated before the replacement hedge counterparties are known.  We call these counterparties anonymous counterparties.  The effects on CVA and MVA will generally be exclusive because MVA largely removes CVA, and CVA is hardly relevant for CCPs.   Effects on KVA and FVA will resemble those on MVA.  We provide a theoretical framework, including anonymous counterparties, and numerical examples.   Pricing XVA by considering counterparties in isolation is inadequate and behaviour must be taken into account.  
\end{abstract}

\section{Introduction}

Valuation adjustments today are typically computed without regard for behaviour, considering each counterparty in isolation \cite{burgard2013funding,green2014kva,green2015mva,kenyon2015warehousing}.  In practice this is not the case, events with respect to one counterparty influence behaviour with respect to other counterparties.  We consider a typical example of a bank hedging a client trade and show that XVA prices can be significantly different when behaviour is taken into account: both higher and lower.   In terms of Accounting, since we price assuming behaviour that is typical of market participants, this fulfulls the requirements of IFRS 13.

\section{Hedging trades with a client}

\begin{figure}
\begin{center}
	\includegraphics[trim=0 20 0 60,clip,width=0.85\textwidth]{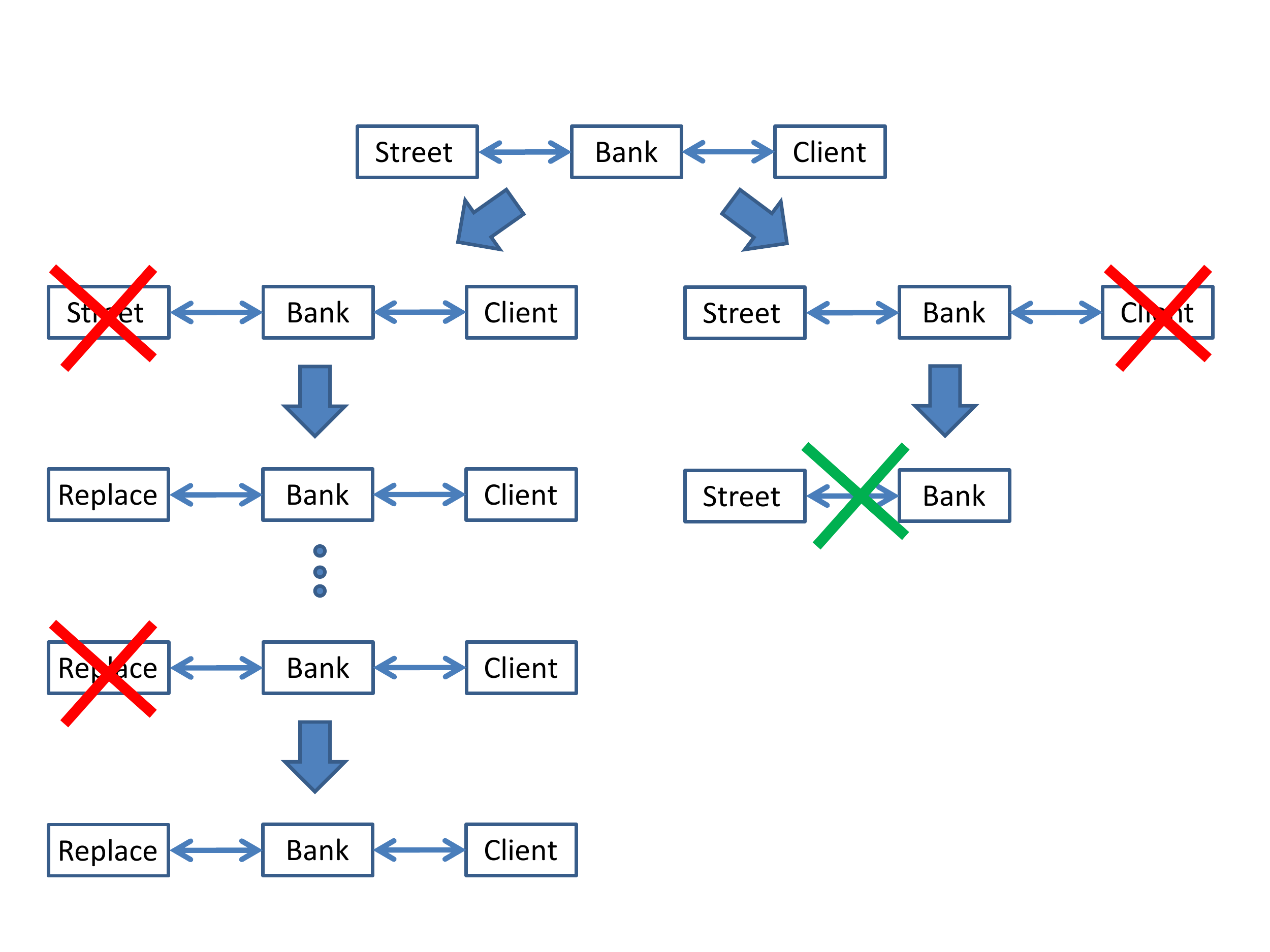}
\end{center}
\caption{Example of behaviour following either the default of a client, or the default of a hedge counterparty.  A hedge is closed (rebalanced) following Client default whereas a hedge is replaced following hedge counterparty default.  If the hedge counterparty again defaults then it will in turn be replaced if the client trade is still active.}
\label{f:events}
\end{figure}

Here we consider a typical example of bank (B) business, providing a trade to a client (C) and buying a hedge from the street (H), see Figure \ref{f:events}.  The hedge with the street will be collateralized and the trade with the client may or may not be collateralized.  

On default of a bank counterparty we observe different behaviour depending on whether the hedge counterparty defaults or the client.  If the client defaults then the hedge is typically closed (Figure \ref{f:events}, RHS).  If the hedge counterparty defaults then the hedge is replaced because the client trade is still alive and the client typically has no interest in closing its trade as it has an economic purpose.  This is a simplification as a hedge may be spread over several counterparties.   But whenever a hedge counterparty defaults then the remaining hedges will be rebalanced.

Considering both sides of the bank's activity and the bank's behaviour radically changes XVA prices.  In this paper we ignore the possibility of changing the hedger dynamically to optimize pricing and XVA.  That is tackled elsewhere (Kenyon and Green, Global Derivatives 2016).

Qualitatively we can explain the effect on CVA as follows.  If H defaults then B has no loss because B has calculated the CVA and charged it to C.  The problem is that C has not defaulted so B must replace the hedge, and this new hedge can default again.  Typically the hedge trades are at mid so B cannot charge the CVA of H to H, contradicting the assumption of independent counterparties that is usually made.  The problem is that the CVA desk may not charge the originating desk for multiple defaults of H and so suffer losses with respect to highly rated clients. The source of loss here for the CVA desk is not the client side but the hedge.   In addition, if the hedge is spread out over several counterparties (with equal default probabilities, and no connection between them) then the expected loss may be increased because there is no netting across counterparties.  However the hedge may net several different clients so the the multi-counterparty-hedge netting effect can be in either direction.  The only way to be sure of effects is to calculate them.  We do not consider any multi-counterparty-hedge netting effect here for simplicity and because weexpect it to be a small fraction of the multi-CVA effect.

When the hedge-side counterparty is replaced, the choice of new hedge counterparty can only be done at that time because any prior choice may itself have defaulted.  Thus we must price into the multi-CVA effect counterparties that are not known at $t=0$.  We call these anonymous counterparties and show that they can be handled relatively simply since we do not care about the identity of the future counterparty but only their riskiness.

Considering MVA we generally have the opposite effect dominating.  That is, the default of the C means that B closes the hedge H so the default probablity driving the MVA cost is from the client side not from the hedge side.  Since hedge counterparties are generally banks,  they will often have a lower hazard rate than clients.  Thus the actual MVA is generally lower than what would be expected from looking at the counterparty in isolation.

The effects on MVA and CVA will generally be exclusive because IM largely removes CVA, and the CVA is low in any case for collateralized counterparties.  Hedge counterparties that do not require IM, e.g. because they are smaller, are attractive because of they do not require IM and so provide cheaper hedges.

Effects on KVA and FVA will resemble those on MVA.  Portfolio rebalancing due to client default will impact KVA and FVA in a similar way to MVA as they are largely exposure profile based.

\section{Methods}

We consider two behavioural cases: firstly a client portfolio hedged with a (riskless) central counterparty (CCP) where the hedge is removed on client default; and secondly a client portfolio hedged with a sequence of defaultable counterparties up to client default.  Given that interest rate portfolios are typical in these settings we model stochastic interest rates.  Our focus is the change in MVA and CVA from the bank's behaviour.

We assume that the default of the bank, client and counterparties are independent for simplicity.  Our development extends \cite{burgard2013funding,green2014kva,green2015mva}.  We price in the risk-neutral measure for simplicity despite the fact that most credit risk is difficult to hedge, see \cite{kenyon2015warehousing} on how to include other measures.  The key technical innovations are 1) including the behaviour of the bank w.r.t. client and hedge default thus breaking the usually-assume independence of counterparties; 2) introducing anonymous counterparties in $t=0$ XVA pricing.  Anonymous counterparties are future hedging counterparties decided in the future according to criteria at $t=0$.  Although unknown at $t=0$ their presence can be included with some assumptions on the future state of the market, i.e. creditworthiness of future hedge counterparties (i.e. CDS spreads) conditional on default of previous hedge counterparties.

Notation is given in Table \ref{t:notation}.

\begin{table}
\begin{tabular}{cl}
\bf Symbol(s) & \bf Meaning \\ \hline
$V, \Vhat_\text{CCP}, V_\text{CCP}$ & riskless price, adjusted CCP price, CCP price \\ 
$\VhatBoth, \VhatC, \VhatH$ & adjusted price total, client-side, and hedge-side \\ 
$\Vhat_\text{to client}$ & adjusted price charged to client \\ 
$P_r,\ P_C,\ P_B,\ P_i$ & bond prices: riskless, client, bank, anonymous \\ 
$r,\ r_C,\ r_B,\ r_i$ & risk-neutral bond drifts: riskless, client, bank, anonymous \\ 
$r_{I;x,y}$ & interest rate on IM posted by $x$ to $y$ \\ 
$r_X$ & interest rate on VM collateral\\ 
$P_\sigma,\ \sigma_C,\ \sigma_B,\ \sigma_i$ & bond volatilities: riskless, client, bank, anonymous \\ 
$R_C,\ R_B,\ R_i$ & recovery rates: client, bank, anonymous \\ 
$J_C,\ J_B,\ J_{\{i\}}$ & default (Poisson) processes:  client, bank, anonymous \\ 
$dW$ & Brownian motion process  \\ 
$\Delta_*$ & differential w.r.t. jump $*$ \\ 
$q_C,\ q_i$ & repo rate on client, anonymous, bonds \\ 
$\lambda_C,\ \lambda_B,\ \lambda_i$ & hazard rates for: client, bank, anonymous \\ 
$X_H$ & variation margin collateral on hedge-side \\ 
$I_{x,y}$ & initial margin posted by $x$ to $y$ \\ 
$\alpha_C,\ \alpha_B,\ \alpha_i$ & hedge-portfolio bond holdings: client, bank, anonymous \\ 
$\phi$, $K$ & use of capital for funding and capital on hedge-side \\ 
$\gamma_K$ & cost of capital (rate) \\ 
$\beta_*$ & cash accounts \\ 
$\Pi_H$ & hedging portfolio to replicate hedge-side \\ 
$\epsilon_H$ & hedging error on bank default on hedge-side \\ 
$g_C,\ g_B,\ g_i$ & closeout on client, bank, and anonymous counterparty default \\ 
$U_H = \VhatH - V_H$ & adjustments value on hedge-side \\ 
$s_* = r_* - r$ & spread over riskless rate for $*$ \\ 
$\tau_{\{i\}}$ & default time of counterparty $i$ \\
$D_*$ & discount factor using discounting $*$ \\ 
$Q$ & rate matrix for counterparty-sequence state transition \\
\hline
\end{tabular}
\caption{Notation.}
\label{t:notation}
\end{table}

\subsection{Client hedged with CCP}

We assume that the CCP is riskless for simplicity in this section.  The next section introduces risky hedge counterparties.  We separate the pricing problem of the adjusted client plus hedge cost $\VhatBoth$ into the two sides linked by behaviour transmitted via client default.
\begin{equation}
\VhatBoth := \VhatC + \VhatH
\end{equation}
where $C$ stands for client and $H$ for hedge.  Note that this is not the separation of (client) pricing as
\begin{equation}
\Vhat_\text{to client} = \VhatC + \VhatH - V
\end{equation}
because the CCP pricing is riskless, i.e. $\Vhat_\text{CCP}=V_\text{CCP}=V\ne\VhatH$, i.e. there is no price adjustment w.r.t. the CCP.  The technical development for $\VhatC$ is standard so we only provide the development for $\VhatH$ to highlight the changes.

The assets available have the following dynamics, where we assume that the bank has only a single class of bonds available for funding for realism (i.e. we assume trading does not drive issuance).  No CCP (hedge $H$) bonds are included as the CCP is assumed riskless, but Client, $C$, bonds are included for the hedge-side price replication as they signal when to close the hedge.
\begin{align}
dP_r/P_r& = r dt + \sigma_r dW\\
dP_C/P_C& = r_C dt + \sigma_C dW - (1 - R_C) dJ_C\\
dP_B/P_B& = r_B dt + \sigma_r dW -  (1 - R_B) dJ_B
\end{align}
We assume zero bond-cds and bond-repo basis so,
\begin{align}
r_C - q_C&= (1-R_C)\lambda_C \\
r_B - r&= (1-R_B)\lambda_B
\end{align}

Initial margin (IM) received cannot be rehypothecated (and CCPs do not post IM) so the self-financing condition only includes IM posted by the bank $I_{x,y}$ where $x,y$ indicates the poster and counterparty respectively:
\begin{align}
\VhatH  - X_H + I_{B,H}  + \alpha_B P_B -\phi K = 0  \label{e:self}
\end{align}
and the cash account changes (prior to rebalancing, but we omit the customary tildes for simplicity) are
\begin{align}
d\beta_r &= - \delta r P_r dt, &  d\beta_C& = -\alpha_C q_C P_C dt, \ \ \ \ \quad d\beta_K = -\gamma_K K dt\\
d\beta_{X} &= -r_X X dt,&  d\beta_{I;x,y}& =  \left(  r_{I;x,y} I_{x,y} - r_{I;y,x} I_{y,x}  \right)dt \qquad x,y\in\{B,H\} 
\end{align}
Using It\^o's lemma we obtain the differential of value of the derivative portfolio towards the CCP and of the replicating portfolio $\Pi_H$:
\begin{align}
d\VhatH + d\Pi_H& = \dVHdt dt +\half \sigma_r^2 P_r^2 \dVVHdPP dt + \dVHdP dP_r + \Delta_B \VhatH dJ_B + \Delta_C \VhatH dJ_C  \label{e:switch} \\
&\quad {}+\delta dP_r + \alpha_B dP_B + \alpha_C dP_C 
- \underbrace{\alpha_C q_C P_C dt}_{d\beta_C} - \underbrace{\delta r P_r dt}_{d\beta_r} \notag \\
&\quad - \underbrace{r_X X dt}_{d\beta_X} - \underbrace{\gamma_K K dt}_{d\beta_K}
-  \underbrace{(r_{I;B,H} I_{B,H} - r_{I;H,B} I_{H,B})dt}_{d\beta_{I;B,H}} \notag
\end{align}
Since this CCP hedge is cancelled on client default (e.g. by trading the opposite position then compressing), all three entities, $B,H,C$  appear in Equation \ref{e:switch}.  So,
\begin{align}
d\VhatH + d\Pi_H& = \left[  \dVHdt  +\half \sigma_r^2 P_r^2 \dVVHdPP + \dVHdP r P_r  - \delta r P_r + \alpha_B r_B P_B  + \alpha_C r_C P_C \right. \notag\\
&\quad \left.{} - \alpha_C q_C P_C  + \delta r P_r - r_X X -\gamma_K K - r_{I;B,H} I_{B,H} \vphantom{\dVVHdPP}\right] dt  \notag\\
&\quad {}+ \left[ \dVHdP \sigma_r P_r +   \delta \sigma_r P_r + \alpha_B \sigma_B P_B + \alpha_C \sigma_C P_C       \right] dW \notag\\
&\quad {}+\underbrace{ \left[\Delta_B \VhatH - \alpha_B(1-R_B) P_B\right]}_{\epsilon_H}  dJ_B \notag\\
&\quad {}+ \left[ \Delta_C \VhatH - \alpha_C(1-R_C) P_C \right] dJ_C  \notag
\end{align}
Setting the $dW,dJ_C$ brackets to zero resolves the values of $\alpha_C$ in the $dt$ bracket (there is no net $\delta$ term in the $dt$ bracket).  The self-financing condition, Equation \ref{e:self} sets the value of $\alpha_B$ so there is no degree of freedom to remove the hedge-side mismatch  $\epsilon_H$ on bank-default (nor on the client side).  Using the self-financing condition for $\alpha_B r_B P_B$ we obtain for the $dt$ bracket, setting collateral received from the CCP to zero (CCPs do not post collateral):
\begin{align*}
 0&=\dVHdt  +\half \sigma_r^2 P_r^2 \dVVHdPP + \dVHdP r P_r  + r_B(-\VhatH  + X_H - I_{B,H}  +\phi K)  \\
 &\quad{}+ \lambda_C ( 1 - R_C) P_C - r_X X -\gamma_K K - r_{I;B,H} I_{B,H}
 \end{align*}
 hence
 \begin{align*}
 \dVHdt  +\half \sigma_r^2 P_r^2 \dVVHdPP + \dVHdP r P_r  - \VhatH(r_B + \lambda_C) &= (r_X-r_B)X+ (r_I-r_B)I_{B,H} -\lambda_C g_C \notag \\
 &\quad {}  +(\gamma_K - r_B\phi)K 
\end{align*}
Now when the client defaults, both the CCP and the bank are non-defaulting so the client closeout with the CCP is at the riskless (CCP) price so
 \begin{align*}
g_C = V_H
\end{align*}
and $V_H$ satisfies the Black-Scholes equation which we can subtract via the anzatz $\VhatH = V_H + U_H$ giving
 \begin{align*}
 \dUHdt  +\half \sigma_r^2 P_r^2 \dUUHdPP + \dUHdP r P_r  - U_H(r_B + \lambda_C) &=   s_B(V_H-X_H-I_H) +s_{X;H} X_H + s_{I;B,H} I_{B,H}\notag \\
 &\quad{} +(\gamma_K - r_B\phi)K
\end{align*}
This gives expressions for the valuation adjustments on the hedge (H, i.e. CCP) side as below using Feynman-Kac.    
\begin{align}
\CVA_\text{CCP} &=  0   \\
\FVA_\text{CCP} &= -\int_0^T  \Ebb\left[  s_B D_{r_B+\lambda_C} (V_H-X_H)    \right]   dt   \\
\ColVA_\text{CCP} &= -\int_0^T  \Ebb\left[  s_X D_{r_B+\lambda_C} X_H    \right]   dt   \\
\MVA_\text{CCP} &= -\int_0^T  \Ebb\left[ (s_B - s_{I;B,H}) D_{r_B+\lambda_C} I_{B,H}    \right]   dt   \\
\KVA_\text{CCP} &= -\int_0^T  \Ebb\left[ (\gamma_K - \phi r_B) D_{r_B+\lambda_C} K_H    \right]   dt   
\end{align}
These will be charged to the client side as they cannot be charged to the CCP, i.e.
\begin{align}
\CVA_\text{to client} &= \CVA_\text{CCP} -  \int_0^T  \Ebb\left[\lgd  \lambda_C D_{r_B,\lambda_C} (V_C- X_C - I_{C,B})^+    \right]   dt   \\
\FVA_\text{to client} &= \FVA_\text{CCP} - \int_0^T  \Ebb\left[  s_B D_{r_B+\lambda_C} (V_C-X_C - I_{B,C})    \right]   dt   \\
\ColVA_\text{to client} &= \ColVA_\text{CCP} - \int_0^T  \Ebb\left[   D_{r_B+\lambda_C} (s_{X;C} X_C + r_{I;C,B} I_{C,B} )  \right]   dt   \\
\MVA_\text{to client} &=\MVA_\text{CCP} - \int_0^T  \Ebb\left[ (s_B - s_{I;B,C}) D_{r_B+\lambda_C} I_{B,C}     \right]   dt   \\
\KVA_\text{to client} &=\KVA_\text{CCP} - \int_0^T  \Ebb\left[ (\gamma_K - \phi r_B) D_{r_F,\lambda_C} K_C    \right]   dt   
\end{align}
That is, the client charge is the sum of the valuation adjustments on the CCP side and on the client side, {\it and} the valuation adjustments on the CCP side incorporate the client default.  This last feature breaks the usual assumption of independence of counterparties for XVA calculations.

\subsection{Client hedged with non-CCP}

When hedging with a non-CCP the hedge counterparty can default, but the bank will replace the hedge for as long as the client is non-defaulting.  Thus the bank will experience periodic costs on hedge defaults.  The hedge-side position is not a single trade (or portfolio) but a sequence of trades with different counterparties.  It is this  hedge-side position that the bank must replicate to price the hedge position.  The  differential of the value of the hedge position is:
\begin{align}
d\VhatH & = \dVHdt dt +\half \sigma_r^2 P_r^2 \dVVHdPP dt + \dVHdP dP_r + \Delta_B \VhatH dJ_B + \Delta_C \VhatH dJC +  \sum_i \Delta_i \Vhati dJ_{\{i\}} \label{e:multiple} 
\end{align}
where $i$ runs over the sequence of hedge counterparties.  We are replicating the cost of the hedge position (actually sequence of positions with different counterparties $i$), not of individual hedge trades with a single counterparty.  This means that only one $\alpha_i$ will be non-zero at any time.
Thus the change in value of the hedge position on the default of a given hedge counterparty must be such that any losses are covered, i.e.
\begin{align}
\Delta_i \Vhati  = (1 - R_i) (V_H - X_H - I_{i,B})^+
\end{align}
Ihe processes $d J_{\{i\}}$ are the {\it sequence} of processes of hedge counterparty defaults.  Apart from the first, the counterparties are anonymous.  When hedging with counterparty $i$ we do not know which counterparty will be next ($i+1$).  This can only be decided at $\tau_{\{i\}}$ when $i$th in the sequence defaults as any prior choice may default prior as well.  

The replicating hedge portfolio now includes (shorted) bonds with the sequence of counterparties $i$ to generate these cashflows on default of each of the $i$, i.e.
\begin{align}
\Pi = \ldots + \sum_i \alpha_i P_i
\end{align}
where the bonds have dynamics, and cash accounts as
\begin{align}
d P_i / P_i = r_i dt + \sigma_i dW - (1-R_i) dJ_i\qquad d\beta_i = -\alpha_i q_i P_i dt
\end{align}
Following a similar development as in the previous section we get to
 \begin{align*}
0= \dVHdt  +\half \sigma_r^2 P_r^2 \dVVHdPP + \dVHdP r P_r  + \lambda_C\Delta_C\VhatH + \sum_i 1_{i} \lambda_i\Delta_i\VhatH - r_X X - \gamma_K K - (r_{I;B,H} I_{B,H} - r_{I;H,B} I_{H,B})
\end{align*}
Note that we do not distinguish between hedge counterparties for IM collateral rates and put $H$ (instead of a sum over $i$ with time periods) for simplicity.  The $1_{i}$ is the indicator function there is a trade with counterparty $i$.  This is a reminder that only one hedge counterparty is used at any point for the position.    

This gives the following expressions for valuation adjustments on the hedge side taking into account multiple hedge counterparty defaults.
\begin{align}
\CVA_H &=  -  \int_0^T  \Ebb\left[ \sum_i 1_{i} s_i D_{r_B+\lambda_C} (V_i- X_i - I_{i,B})^+    \right]   dt   \label{e:cva} \\
\FVA_H &=  - \int_0^T  \Ebb\left[  s_B D_{r_B+\lambda_C} \sum_i 1_{i} (V_H-X_H - I_{B,i})    \right]   dt   \\
\ColVA_H &= - \int_0^T  \Ebb\left[   D_{r_B+\lambda_C}\sum_i 1_{i} (s_{X;C} X_C + r_{I;i,B} I_{i,B} )  \right]   dt   \\
\MVA_H &= - \int_0^T  \Ebb\left[ s_{I;B,i} D_{r_B+\lambda_C}\sum_i 1_{i} I_{B,i}     \right]   dt   \\
\KVA_H &= - \int_0^T  \Ebb\left[ (\gamma_K - \phi r_B) D_{r_B+\lambda_C} \sum_i 1_{i}K_i    \right]   \label{e:kva} dt   
\end{align}
These expectations appear challenging to compute because of the $1_{i}$ term which changes them from the usual one dimensional expression to (theoretically) infinitely dimensional.  There is one dimension per hedging counterparty.  In addition there is the implicit assumption that the CDS spreads of the sequence of counterparties are known, even if the actual counterparties are not known.  

However, the situation is less challenging than it appears, consider Equation \ref{e:cva} where we combine elements not affected by the CDS level, or replacement, into $f(t)$
\begin{align}
\int_0^T  \Ebb\left[ \sum_i 1_{i} s_i f(t)   \right]  dt   \label{e:1} 
\end{align}
We know that exactly one of the $1_{i}$ is non-zero at any time, so if all the $s_i$ are equal (for whatever reason), to say $s_1$, we have
\begin{align}
\int_0^T  \Ebb\left[ \sum_i 1_{i} s_i f(t)   \right]  dt    = \int_0^T  \Ebb\left[   s_1 f(t)   \right]  dt   \label{e:2}
\end{align}
Now suppose instead that prior to any defaults we have  $s_i$ all different, so we now need to consider how many defaults have occured up to $t$ to know which $s_i$ is being used.  Thus the state transitions between hedge-side counterparties form a semi-Markov process with rate matrix $Q$:
\begin{equation}
Q =
\begin{pmatrix}
-\lambda_1 & \lambda_1 & 0 & 0 & ...  & 0 & 0\\
0 & -\lambda_2 & \lambda_2 & 0 & ...  & 0 & 0 \\
\vdots &           &                 &     & \ddots & & \\
0 & 0 & 0 & 0 & ... & -\lambda_n & \lambda_n \\
0 & 0 & 0 & 0 & ... & 0 & 0 \\
\end{pmatrix}
\end{equation}
The full $Q$ is infinite but can be truncated at $n$ transitions without material error where $n$ can be calculated from the $\lambda_i$ for a given required accuracy.  The time zero state vector is simply $\{1,0,\ldots,0 \}$ expressing that there is one hedge-side counterparty at the start and this is the first one.  

We can now express Equation \ref{e:1} fully generally as:
\begin{align}
\int_0^T  \Ebb\left[ \sum_i 1_{i} s_i f(t)   \right]  dt    = \int_0^T  \Ebb\left[ \sum_i\left( \text{PDF}_\text{SMP}(\{1,0,\ldots,0 \},Q,t,i) s_i \right) f(t)   \right]  dt   \label{e:2}\\
\end{align}
Where $ \text{PDF}_\text{SMP}(\{1,0,\ldots,0 \},Q,t)$ is the PDF of the  semi-Markov process defined by the given starting vector, the rate matrix $Q$ for state $i$, i.e. the i-th hedge-side counterparty.  This can be calculated efficiently by matrix exponentiation.

In summary, in both hedge cases the hazard rate of the hedge does not appear in the discounting of the valuation adjustments, but the hazard rate of the client does.  Thus, in practice, usual hedging behaviour breaks the assumption of independence of counterparties.

\section{Numerical Examples}

Here we consider a client trade hedged with the street.  We consider street counterparties directly and via a central counterparty (CCP).  However, we do not compute the effect of multiple CCP defaults since even a single CCP default is likely to be market-changing beyond what is captured in our equations.  This means that trading via a CCP is relevant for MVA but not the other XVA.  Althougth this is only one of the XVA, the effect of behaviour on MVA with a CCP is highly significant precisely because of the low CCP default probability relative to clients of the bank.  Effects on MVA and CVA will generally be exclusive as MVA largely removes CVA.

We assume that hedge trades are at mid, i.e. the hedging counterparty does not charge the bank for the hedger's MVA and vice versa.  

We consider exposures from interest rate positions.  Looking at Equations \ref{e:cva}---\ref{e:kva} we note that if earlier defaults do not change subsequent  then multiple defaults simply change the default probabilty, not the quantities involved.  For non-credit portfolios this is the case assuming the usual independence of expsoure and default, i.e. no wrong-way risk and no right-way risk.

Equations \ref{e:cva}---\ref{e:kva} are challenging to compute directly, but given that they converge quickly for multiple defaults, we make the following approximations:
\begin{itemize}
\item include only three possible defaults (not just the first default as usual).  For the cases considered, i.e. maximum CDS spread of 5\%\ three defaults up to 30Y captures 93\%\ of events, and for 2.5\%\ this captures 99\%\ of events.   Missed event will have a very small impact as they occur late in the time considerd, i.e. missing 7\%\ of events does not have a 7\%\ impact but much less than 5\%.
\item exposure profiles assumed from non-credit products and assume that these are unaffected by previous defaults.  We consider three exposure profiles, either from MVA or CVA depending on the case
\begin{itemize}
\item  Triangular decreasing to zero at maturity
\item Flat
\item Triangular increasing from zero at start

Since we consider percentage changes in MVA and CVA, these  profile specifications are sufficient.  We also consider two cases for CVA effects
\end{itemize}
\item assume that all available hedge counterparties have the same undisturbed hazard rate
\item assume that each default increases available hedge counterparty hazard rates by a fixed multiplier, 20\%.  
\end{itemize}
We report results in terms of the changes on the hedge side as the client side is, by assumption, unaffected. The effects may be slightly underestimated in practice because hazard rates of hedge counterparties are assumed independent of non-hedge entity defaults.  That is, only direct contaigon is included, and at a relatively mild level (20\%\ increase in CDS spread).

Table \ref{t:mvaccp} shows the relative effect on MVA when hedging client trades through a CCP, i.e. a hedge counterparty that is considered default free.  The fact that hedges will be closed on client default means that just looking at the hedge counterparty (where the MVA occurs) is misleading.  For reasonably risky clients (250bpds CDS spread) decreases of MVA of roughly 20\%\ occur for flat and increasing profiles at 10Y portfolio length.  For 30Y portfolios there are vey significant decreases in MVA, 30\%\ to 50\%\ for decreasing-to-increasing MVA profiles.  For risky counterparties (500bps) there are major decreases in MVA for 10Y portfolios and above.

Trades with a CCP combine many different counterparties so the actual decrease in MVA will be a mixture of the different client CDS results shown.

The results for MVA with a CCP are also valid, exactly as they are, for non-CCPs where hedge counterparties are replaced when they default.

Table \ref{t:cva2} and Table \ref{t:cva3} shows the change in CVA from the combination of two effects: hedge closure on client default; and hedge replacement on hedge default.  These effects act in opposite directions and we consider ranges of client CDS level and hedge CDS level for different portfolio lengths  from 5Y to 30Y.  Table \ref{t:cva2} shows results assuming that the next hedge CDS level available on previous hedge default is the same level.  This is a rather aggresive position as bank defaults tend to affect each other.  Accordingly Table \ref{t:cva3} shows the changes in CVA where we assume that the available hedge CDS level increases by 20\%\ on each previous hedge default.

Table \ref{t:cva2} shows CVA changes where client CDS and hedge CDS effects become more asymmetric as the length of the portfolio considered increases.  The asymmetry comes from the increased possibility of multiple hedge replacements (multiple CVA charges) on the hedge-side.  The symmetry of the effects is modified by the portfolio's exposure profile.  Decreasing profiles emphasising client default and increasing profiles emphasising hedge-replacement.  For  long, 30Y, trades highly significant ($>$50\%) changes in CVA on the hedge-side can occur in both directions.  However, for typical decreasing exposure portfolios and riskier clients than hedges there are 10\%\ to 40\%\ decreases in hedge-side CVA.  Of course, for incremental trades the incremental exposure profile needs to be considered (as modified by any collateralization).

Table \ref{t:cva3} shows CVA changes where we assume that the available hedge CDS level increases by 20\%\ on each previous hedge default.  On default of a major bank, increases in remaining-bank CDS spreads have occured in the past.  However, with major banks we can expect MVA to be in place so this table applies to hedging with smaller banks or legacy portfolios not covered by MVA.  With this mild level of CDS spread increase, 20\%, increased CVA is evident.  For the case of a low-CDS client, e.g. a Special Purpose Vehicle where the bank can look through to underlying safe assets, the increases in CVA on the hedge-side from multiple hedge defaults-and-replacement can be more than double at the extreme.   This effect is non-linear in the hedge-side CDS spread, appearing much more for hedege-side CDS spreads of 250bps than for 100bps.

\begin{table}
\begin{centering}
\begin{tabular}{cccc}
Client CDS& \multicolumn{3}{c}{MVA Profile}\\
 \text{100. bps} & \text{Decreasing} & \text{Flat} & \text{Increasing} \\ \hline
 \text{1Y} & \text{-1$\%$} & \text{-1$\%$} & \text{-1$\%$} \\
 \text{5Y} & \text{-3$\%$} & \text{-4$\%$} & \text{-5$\%$} \\
 \text{10Y} & \text{-5$\%$} & \text{-8$\%$} & \text{-10$\%$} \\
 \text{30Y} & \text{-13$\%$} & \text{-18$\%$} & \text{-26$\%$} \\ \hline
 \text{250. bps} & \text{Decreasing} & \text{Flat} & \text{Increasing} \\ \hline
 \text{1Y} & \text{-1$\%$} & \text{-2$\%$} & \text{-3$\%$} \\
 \text{5Y} & \text{-6$\%$} & \text{-10$\%$} & \text{-13$\%$} \\
 \text{10Y} & \text{-12$\%$} & \text{-17$\%$} & \text{-23$\%$} \\
 \text{30Y} & \text{-28$\%$} & \text{-38$\%$} & \text{-51$\%$} \\ \hline
 \text{500. bps} & \text{Decreasing} & \text{Flat} & \text{Increasing} \\\hline
 \text{1Y} & \text{-3$\%$} & \text{-4$\%$} & \text{-5$\%$} \\
 \text{5Y} & \text{-12$\%$} & \text{-18$\%$} & \text{-24$\%$} \\
 \text{10Y} & \text{-22$\%$} & \text{-31$\%$} & \text{-41$\%$} \\
 \text{30Y} & \text{-44$\%$} & \text{-57$\%$} & \text{-74$\%$} \\\hline
\end{tabular}
\caption{Relative percentage decrease in MVA on CCP side from closing hedge with CCP on client default.  Table shows effect of MVA profiles and portfolio maturities (in years) for different client CDS levels (in basis points).  Profiles: decreasing = triangular decreasing to zero at maturity; flat = flat; increasing = triangular increasing from zero.  Bank CDS is 100bps in all cases.  CCP is treated as riskless.}
\label{t:mvaccp}
\end{centering}
\end{table}

\begin{table}
\begin{centering}
\begin{tabular}{ccccc|cccc|cccc}
  \multicolumn{13}{c}{\bf No jump in next-hedge CDS spread on each previous-hedge default}\\
 & \multicolumn{4}{c}{Decreasing CVA Profile} &  \multicolumn{4}{c}{Flat CVA Profile} &   \multicolumn{4}{c}{Increasing CVA Profile} \\
 & &  \multicolumn{3}{c}{Hedge CDS} &  & \multicolumn{3}{c}{Hedge CDS} &  & \multicolumn{3}{c}{Hedge CDS} \\
 &  \multicolumn{12}{c}{\bf 5Y}\\
 &\text{\bf bps} & 50. & 100. & 250. & \text{\bf bps} & 50. & 100. & 250. & \text{\bf bps} & 50. & 100. & 250. \\
  \parbox[t]{2mm}{\multirow{9}{*}{\rotatebox[origin=c]{90}{Client CDS}}}
 &50. & \text{0$\%$} & \text{1$\%$} & \text{5$\%$} & 50. & \text{0$\%$} & \text{2$\%$} & \text{8$\%$} & 50. & \text{0$\%$} & \text{3$\%$} & \text{11$\%$} \\
 &100. & \text{-1$\%$} & \text{0$\%$} & \text{4$\%$} & 100. & \text{-2$\%$} & \text{0$\%$} & \text{6$\%$} & 100. & \text{-3$\%$} & \text{0$\%$} & \text{8$\%$} \\
 &250. & \text{-5$\%$} & \text{-4$\%$} & \text{0$\%$} & 250. & \text{-8$\%$} & \text{-6$\%$} & \text{0$\%$} & 250. & \text{-10$\%$} & \text{-8$\%$} & \text{0$\%$} \\
 &500. & \text{-11$\%$} & \text{-10$\%$} & \text{-6$\%$} & 500. & \text{-16$\%$} & \text{-14$\%$} & \text{-9$\%$} & 500. & \text{-22$\%$} & \text{-19$\%$} & \text{-12$\%$} \\
  &  \multicolumn{12}{c}{\bf 30Y}\\
& 50. & \text{0$\%$} & \text{7$\%$} & \text{29$\%$} & 50. & \text{0$\%$} & \text{10$\%$} & \text{45$\%$} & 50. & \text{0$\%$} & \text{16$\%$} & \text{76$\%$} \\
& 100. & \text{-7$\%$} & \text{0$\%$} & \text{20$\%$} & 100. & \text{-9$\%$} & \text{0$\%$} & \text{31$\%$} & 100. & \text{-14$\%$} & \text{0$\%$} & \text{52$\%$} \\
& 250. & \text{-22$\%$} & \text{-17$\%$} & \text{0$\%$} & 250. & \text{-31$\%$} & \text{-24$\%$} & \text{0$\%$} & 250. & \text{-43$\%$} & \text{-34$\%$} & \text{0$\%$} \\
& 500. & \text{-40$\%$} & \text{-36$\%$} & \text{-23$\%$} & 500. & \text{-52$\%$} & \text{-47$\%$} & \text{-31$\%$} & 500. & \text{-69$\%$} & \text{-65$\%$} & \text{-46$\%$} \\

\end{tabular}
\caption{Relative percentage change in CVA on hedge-side from combined hedge replacement on hedge default and hedge closure on client default.  Table shows effect of CVA profiles and portfolio maturities (in years) for different hedge-side CDS levels and client CDS levels (in basis points).  Profiles: decreasing = triangular decreasing to zero at maturity; flat = flat; increasing = triangular increasing from zero.  Bank CDS is 100bps in all cases.}
\label{t:cva2}
\end{centering}
\end{table}

\begin{table}
\begin{centering}
\begin{tabular}{ccccc|cccc|cccc}
  \multicolumn{13}{c}{\bf 20\%\ jump in next-hedge CDS spread on each previous-hedge default}\\
 & \multicolumn{4}{c}{Decreasing CVA Profile} &  \multicolumn{4}{c}{Flat CVA Profile} &   \multicolumn{4}{c}{Increasing CVA Profile} \\
 & &  \multicolumn{3}{c}{Hedge CDS} &  & \multicolumn{3}{c}{Hedge CDS} &  & \multicolumn{3}{c}{Hedge CDS} \\
 &  \multicolumn{12}{c}{\bf 5Y}\\
 &\text{\bf bps} & 50. & 100. & 250. & \text{\bf bps} & 50. & 100. & 250. & \text{\bf bps} & 50. & 100. & 250. \\
  \parbox[t]{2mm}{\multirow{9}{*}{\rotatebox[origin=c]{90}{Client CDS}}}
& 50. & \text{0$\%$} & \text{2$\%$} & \text{7$\%$} & 50. & \text{0$\%$} & \text{3$\%$} & \text{11$\%$} & 50. & \text{1$\%$} & \text{4$\%$} & \text{15$\%$} \\
& 100. & \text{-1$\%$} & \text{1$\%$} & \text{5$\%$} & 100. & \text{-2$\%$} & \text{1$\%$} & \text{8$\%$} & 100. & \text{-2$\%$} & \text{1$\%$} & \text{12$\%$} \\
& 250. & \text{-5$\%$} & \text{-3$\%$} & \text{1$\%$} & 250. & \text{-7$\%$} & \text{-5$\%$} & \text{2$\%$} & 250. & \text{-10$\%$} & \text{-7$\%$} & \text{3$\%$} \\
& 500. & \text{-11$\%$} & \text{-9$\%$} & \text{-5$\%$} & 500. & \text{-16$\%$} & \text{-14$\%$} & \text{-7$\%$} & 500. & \text{-21$\%$} & \text{-18$\%$} & \text{-10$\%$} \\
  &  \multicolumn{12}{c}{\bf 30Y}\\
& 50. & \text{1$\%$} & \text{10$\%$} & \text{39$\%$} & 50. & \text{2$\%$} & \text{15$\%$} & \text{62$\%$} & 50. & \text{3$\%$} & \text{23$\%$} & \text{109$\%$} \\
& 100. & \text{-5$\%$} & \text{3$\%$} & \text{29$\%$} & 100. & \text{-8$\%$} & \text{4$\%$} & \text{46$\%$} & 100. & \text{-11$\%$} & \text{6$\%$} & \text{79$\%$} \\
& 250. & \text{-21$\%$} & \text{-15$\%$} & \text{7$\%$} & 250. & \text{-30$\%$} & \text{-21$\%$} & \text{10$\%$} & 250. & \text{-42$\%$} & \text{-30$\%$} & \text{16$\%$} \\
& 500. & \text{-40$\%$} & \text{-35$\%$} & \text{-19$\%$} & 500. & \text{-52$\%$} & \text{-46$\%$} & \text{-26$\%$} & 500. & \text{-69$\%$} & \text{-63$\%$} & \text{-39$\%$} \\
\end{tabular}
\caption{Changes in CVA on hedge-side as in Table \ref{t:cva2} but now we assume that the next hedge CDS level is increased by 20\%\ on each previous-hedge default.  We assume that the client is from a different sector so the client CDS is unchanged by hedge-side defaults.  Note that the very approximate symmetry present in Table \ref{t:cva2} is reduced. }
\label{t:cva3}
\end{centering}
\end{table}

\section{Conclusions}

Considering actual bank behaviour in XVA significantly decreases client prices from MVA on the hedge side.  Prices changes to the client from CVA on the hedge side, with hedge counterparties where there is no IM or legacy portfolios, show significant increases and decreases depending on the relative CDS levels of client and hedge.  However, the main effect for CVA is any increase in available hedge CDS levels caused by the default of previous hedge counterparties.  This leads to significant increases in CVA on the hedge-side in most cases considerd.    Pricing XVA by considering counterparties independently of each other is inadequate for both MVA on the hedge side and CVA where there is no IM, bank behaviour needs to be included.  Effects on KVA and FVA will resemble those on MVA.  Portfolio rebalancing due to client default will impact KVA and FVA in a similar way to MVA as they are largely exposure profile based, so behaviour needs including for all XVAs.

\bibliographystyle{chicago} 
\bibliography{XVAbibliography}

\end{document}